# BOUNDARY CONDITIONS IN DIFFUSIONAL GROWTH AND SEDIMENTATION

Vladimir Privman and Jongsoon Park

Department of Physics and Center for Advanced
Materials Processing, Clarkson University
Potsdam, New York 13699–5820, USA

## INTRODUCTION

In many applications, transport of particles can be described by the diffusion equation, or its convective-diffusion generalizations, in part of three-dimensional space. In particular, in surface deposition or in growth of aggregates or sediments, particles in the suspension outside the surface or aggregate boundary can be considered as diffusing under the influence of applied forces, such as the double-layer interactions or gravity. However, the diffusion equation cannot be used *at* the substrate or *inside* the aggregate. One convenient and widely used approach has been to impose boundary conditions at the surface, to supplement and completely define the diffusion problem outside it.[1-11] This short survey describes some theoretical results on the use of such boundary conditions and their physical interpretation.

In order to define the geometry of the problem, let us consider Figure 1 (next page). We will assume that the surface is planar, at $x = 0$, and for simplicity use the one-dimensional version of the diffusion equation whenever this causes no confusion. Thus, for $x > 0$, the coordinate $x$ measures the distance away from the substrate or growing aggregate. Some important problems lack planar symmetry. For instance, in the growth of colloid particles from small sub-units, the aggregates are typically spherical, and the radial diffusion equation must be used.

Here, we assume that suspended particles at $x > 0$ move according to the diffusion equation:

$$\frac{\partial c}{\partial t} = D_0 \frac{\partial}{\partial x}\left(\frac{\partial c}{\partial x} + c\frac{dV}{dx}\right) \qquad (1)$$

where $D_0$ denotes the particle diffusion constant and the subscript 0 refers to the dilute-suspension limit. The suspended particle number concentration is $c$, so that for spherical particles, for instance, of radius $r$, the volume filling fraction $\phi$ is

$$\phi = 4\pi r^3 c/3 \qquad (2)$$

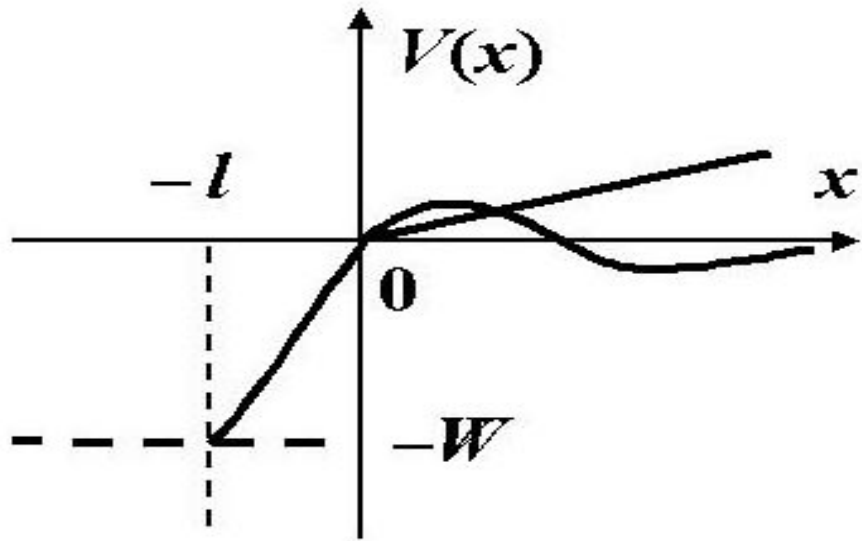

**Figure 1.** Schematic representation of the dimensionless potential energy $V(x)$ with the typical double-layer and linear forms shown.

It is assumed that for $x > 0$, particles move in the potential energy $U(x)$, here only a function of $x$ to ensure planar symmetry. We define the dimensionless potential energy as $V \equiv U/kT$. The potential energy can be double-layer, for colloid growth, or linear, as in sedimentation; see Figure 1. The dilute-suspension diffusion equation can be used[12] approximately up to $\phi \simeq 0.1$, for particles with relatively short-range interactions as compared to their size, so that they can be regarded similar to hard spheres. Note that some of the longer-range interactions of a given particle with other particles, especially with those in the aggregate, will be included phenomenologically as a contribution to the effective potential energy $U(x)$. The van der Waals dispersion-force component of the double-layer potential energy is one such contribution.

A diffusional description applies for many problems. Specifically, in monolayer surface deposition (or detachment) of particles from solution,[4,7,8] the substrate is fixed at $x = 0$. In other cases, such as growth of submicron colloid particles from nano-size sub-units,[1,9,10] the non-planar, typically spherical, surface actually moves in time. In many instances this motion of the growing surface can be treated as a slow process as compared to particle diffusion. In such cases an approximation can be used whereby the diffusion equation is solved with the fixed boundary at $x = 0$. The solution is then used to calculate the flux of matter to the surface, or away from it in cases of detachment/dissolution.[1,4,7,9,10] The slow motion of the surface itself is then treated separately. Furthermore, particle motion in a dilute solution is much better accessible to experimental probes and understood theoretically than the dynamics of a concentrated system. Therefore it is frequently convenient to lump the effects of the surface and growing, restructuring aggregate in a few-parameter phenomenological boundary condition at $x = 0$.

A typical approach to such problems has been to use boundary conditions that impose a linear relation between the concentration $c$ and its derivatives,

$$c = \alpha \frac{\partial c}{\partial x} + \beta + \gamma \frac{\partial c}{\partial t} \qquad \text{(at } x = 0\text{)} \qquad (3)$$

The term on the left-hand side is always there. The three terms on the right-hand side vary in their importance depending on the problem at hand. In the next section, we will survey the various forms typically assumed and describe in detail the choice appropriate for growth and sedimentation. In the last section, we consider aspects of diffusion in concentrated suspensions. Finally, when there is no planar symmetry, the coordinate-derivative term in (3) becomes $\alpha\,\hat{n}\cdot\vec{\nabla}c$, where $\hat{n}$ is the unit vector orthogonal to the surface, pointing into the dispersion.

## BOUNDARY CONDITIONS FOR DIFFUSION

The relative importance of the terms on the right-hand side of Eq. (3) depends on the problem at hand. Let us consider the well known example of monolayer surface deposition and detachment.[8] In this case, the boundary condition is usually written as:

$$\frac{\partial c}{\partial t} = -\kappa c + k \qquad (\text{at } x=0) \tag{4}$$

The first term describes the depletion of the diffusing-particle number at $x=0$ owing to their adhesion to the surface. The second term corresponds to particle detachment. Within a Langmuirian approach,[8] we take:

$$\kappa = \lambda\left(\theta_{\max} - \theta\right) \qquad \text{and} \qquad k = \omega\theta \tag{5}$$

where $\theta$ is the fractional surface coverage and $\theta_{\max}$ is the maximum coverage, while $\lambda$ and $\omega$ are constants. At equilibrium, when the time-derivative vanishes, the above equations yield the famous Langmuir-isotherm relation between the surface coverage and particle concentration near the surface, and, in the absence of potentials, in the bulk of the solution: $c \propto \theta/\left(\theta_{\max}-\theta\right)$.

The above example illustrates that the coefficients $\alpha, \beta, \gamma$ in Eq. (3) can be time-dependent coupled dynamical variables. In another case, that of chemical reactions,[2,3,6] the appropriate relation describes partial reaction probability. It has been established[2,3,6] that the "radiation boundary condition" is applicable: $c$ is proportional to the normal derivative at the reaction radius. In our notation this would correspond to only the coefficient $\alpha$ surviving on the right-hand side of Eq. (3), while $\beta, \gamma = 0$.

In recent work on diffusional growth of spherical colloid aggregates, we explored the form of the boundary condition for the case when the surface is not rigid but rather is itself dynamically evolving.[10] This study was extended to droplet-shaped objects growing on two-dimensional substrates.[11] Here, we survey this approach and offer a different derivation, which is less complicated (at the expense of being somewhat less definitive) but is also more generally applicable. We argue that similar considerations can be applied to sedimentation of submicron or larger suspended particles, provided the growth of the sedimented "cake" layer can be regarded as slow, i.e., the gravitational (centrifugal) acceleration is not too large.

Consider an aggregate or cake-layer slowly growing or dissolving by capture or release of singlet units to/from solution. These "building block" singlets can be solute species or larger suspended particles, depending on the system. Typically in such growth there is no well-defined sharp surface. Instead, there is a region for $x < 0$ in which the particle motion changes from simple dilute-suspension diffusion to whatever restructuring and relaxation dynamical processes are going on deep inside the growing structure. In many experimental situations these restructuring processes are quite fast, leading to dense rather than fractal structures.[9] However, these processes are complicated and cannot be modeled by extensions of the diffusion equation. Still, we assume that for some depth $-\ell < x < 0$, the dilute-limit diffusion Eq. (1) does apply. Such an approximate description deteriorates fast past $x = -\ell$; see Figure 1. The effective potential energy seen by the particles in this region reflects the onset of them being bound within the growing structure and the corresponding increase in the particle concentration. Thus, as shown in Figure 1, we assume that $V(x)$ drops to a large negative value $-W$ at $x = -\ell$, while the particle concentration reaches a large value (as compared to the concentration in the dilute suspension outside, at $x > 0$), denoted $c_W$.

There is evidence in the literature[12] that the dilute diffusion description applies up to volume filling fractions $\phi_W \simeq 0.1$, for colloid (submicron size) and larger particles which can be approximately treated as hard spheres. We assume that $c_W$ and the potential energy value $U(-\ell) = -kTW$ are properties of the structure of the outer layer of the growing cake and are not dependent on the initial concentration or on time, the latter after a short transient time interval. Furthermore, we assume that the concentration within $-\ell \leq x \leq 0$ is in thermodynamic equilibrium. These assumptions suggest that the values of the concentration at $x = 0$ and $x = -\ell$ are related as follows:

$$c(0) = c_W e^{-W} \tag{6}$$

The above arguments suggest that the boundary condition in this case corresponds to $\alpha, \gamma = 0$ in Eq. (3), and to fixed $\beta = c_W e^{-W}$, i.e., to:

$$c = \beta \left(= c_W e^{-W}\right) \qquad \text{(at } x = 0\text{)} \tag{7}$$

A somewhat more detailed study,[10] not presented here, shows that while $\beta$ is a product of one large parameter, $c_W$, and one small parameter, $e^{-W}$, the coefficient $\alpha$ is a product of two small parameters, $\ell$ and $e^{-W}$, and that the omission of the time-derivative term $\propto \gamma$ changes only the initial transient behavior of the concentration near $x = 0$. The resulting boundary condition, $c(0) = \beta$, is thus surprisingly simple and involves only a single phenomenological parameter, the value of the concentration at the growing or dissolving aggregate or cake surface. For a completely irreversible capture of particles by the growing aggregate, the effective energy barrier $W$ is very large and $\beta \to 0$, recovering the celebrated boundary condition of Smoluchowski.[1]

## CONCENTRATED SUSPENSIONS

One might ask how far can the present approach be taken into the region of the concentrated particle system, $x < 0$, and whether corrections are needed due to nonzero concentration for $-\ell < x < 0$. In order to address this matter, let us survey certain aspects of diffusion in concentrated dispersions.[12–14] Some relations of the ordinary diffusion theory remain unchanged when the particle system is no longer dilute. For instance, if $\vec{v}$ represents the average particle velocity, then the particle number flux $\vec{J}$ is:

$$\vec{J} = c\vec{v} \tag{8}$$

using three-dimensional vector notation. However, the contribution $\vec{J}_F$ to flux, caused by a force field $\vec{F}$, can only be approximately assumed proportional to that force:

$$\vec{J}_F = -c\vec{F}/f(c) \tag{9}$$

Strictly speaking, this relation is only exact in the dilute limit, $c \to 0$, when the friction constant has the value $f_0$. For denser systems, the relation need no longer be local, and the friction constant $f(c)$, which is the inverse of the mobility, is defined only approximately.

For denser systems, some relations become generally nonlocal and include memory, in that they involve concentration and possibly other properties at nearby points and at earlier times. Perhaps the most familiar such relation has been between the concentration gradient of the suspended particles and their number flux.[13] The dilute-limit-like relation

$$\vec{J} = -D(c)\vec{\nabla}c \qquad (\text{for } V(\vec{R}) = 0) \tag{10}$$

only becomes exact as $c \to 0$ and $D(c) \to D_0$. A more general nonlocal relation with memory is:

$$\vec{J}(\vec{R},t) = -\int d^3\vec{\rho} \int_0^\infty d\tau\, \mathcal{D}(\vec{\rho},\tau)\, \vec{\nabla}c\left(\vec{R}+\vec{\rho},\, t-\tau\right) \tag{11}$$

where the kernel $\mathcal{D}(\vec{\rho},\tau)$ could be a tensor in a more general case. Since relation (11) can be "deconvoluted" to become a product in the Fourier-transformed representation, it can be directly probed by light scattering measurements in some systems.[13,14]

The following two relations remain valid for any concentration:

$$\frac{\partial c}{\partial t} = -\vec{\nabla}\cdot\vec{J} \tag{12}$$

$$\vec{F}_{\text{appl}} = -kT\, \vec{\nabla}V(\vec{R}) \tag{13}$$

Here, $\vec{F}_{\text{appl}}$ is the applied force per particle. However, the expression for the effective thermodynamic driving force per particle, $\vec{F}_{\text{grad}}$, resulting from the particle concentration gradient, again can be only written as an approximate generalization of the dilute-limit expression. Assuming that we can define the osmotic pressure, $\Pi(c)$, for the particles in the suspension, and that local equilibrium applies so that $\Pi(c)$ is a thermodynamic function of $c$ (and of the temperature $T$, which is assumed constant throughout this work), we have the relation:

$$\vec{F}_{\text{grad}} = -\frac{1}{c}\left(\frac{d\Pi}{dc}\right)\vec{\nabla}c \tag{14}$$

More general formal relations can be written, with this force proportional to the gradient of the chemical potential, accounting for both direct and hydrodynamic particle-particle interactions, as well as applied forces such as gravity.

How far into the region $x < 0$ (see Figure 1) can the approximate Eqs. (9), (10), (14) be used? For hard-sphere-like particle suspensions, the literature[12] seems to suggest that these relations will break down for volume fractions $\phi$, see Eq. (2), somewhere between 0.2 and 0.4. There are two common applications of such relations. Firstly, the sedimentation-velocity ratio $K(c)$ can be defined:

$$K(c) \equiv f_0/f(c) \qquad (15)$$

and the drag-force relation $\vec{F} = -f_0\vec{v}/K(c)$ probed directly to compare $K(c)$ values to theoretical small-$c$ expansion approximations.[12]

Secondly, all the above relations can be combined to yield a variant of the diffusion equation, with the expression for the flux obtained by adding the external-force and gradient fluxes:

$$\frac{\partial c}{\partial t} = \vec{\nabla} \cdot \left[ D(c)\vec{\nabla}c + \frac{kT}{f(c)}c\vec{\nabla}V \right] \qquad (16)$$

and the generalized Einstein relation for the gradient-diffusion coefficient:

$$D(c) = \frac{1}{f(c)} \left( \frac{d\Pi}{dc} \right) \qquad (17)$$

The latter relation reduces to the familiar $D_0 = kT/f_0$ in the limit $c \to 0$, where we used the ideal equation of state $\Pi = kTc$. The above relations can be used as good approximations for $c > 0$ but with volume fractions below the values at which solid-formation or glassy dynamical effects set in. The diffusion Eq. (16) requires a virial expansion for the osmotic pressure $\Pi(c)$, to derive approximations[12] valid for small positive $c$.

We assumed that the dynamics of the aggregating system is slow, so that the surface region of the aggregate is in equilibrium. In order to derive the boundary condition at the surface, we used the equilibrium expression $c \propto e^{-V(x)}$ for small negative $x$. We further assumed that both the effective potential $U(x) = kTV(x)$ felt by a particle and the concentration near $x = -\ell$ do not depend on the concentration profile outside, at $x > 0$, but rather are determined by the structure of the surface of the aggregate. Let us point our that particle-particle (singlet-singlet) interaction effects are accounted for in two different ways in phenomenological relations like Eq. (16). Firstly, short-range interactions are included in thermodynamic properties, such as $\Pi(c)$, or dynamical properties for slow, nearly-steady-state dynamics, such as $K(c)$. Secondly, longer-range interactions contribute to the effective potential $U(\vec{R})$, which also includes the external forces. Avoiding double-counting is therefore a challenge in setting up approximations of this sort.[12–14]

We note that for the modified diffusion-like description given by the equations summarized earlier in this section, the equilibrium assumption of slow dynamics for $x < 0$ can be expressed by the approximate flux balance, equivalent to the force balance $\vec{F}_{\text{grad}} = \vec{F}_{\text{appl}}$, which yields the approximate equation for the (time-independent) concentration, assuming the planar symmetry:

$$\frac{1}{c}\left(\frac{d\Pi}{dc}\right)\frac{dc}{dx} = kT\frac{dV}{dx} \qquad (18)$$

In the dilute limit, with $\Pi = kTc$, this relation obviously integrates to $c \propto e^{-V(x)}$. Otherwise, numerical integration will be needed, with a virial expansion for $\Pi(c)$ used for obtaining approximations for small but nonzero $c$. The latter depend strongly on the details of the particle-particle interactions.

However, no matter the specifics, the modification represented by Eq. (18) does not actually change our basic conclusion. Indeed, it is obvious from its form that Eq. (18) can be integrated from some reference negative $x$ value, $-\ell$, to $x = 0$, and while the expression for $\beta$ in Eqs. (6)-(7) may change, the form of the boundary condition will not: it will still be without the derivative terms entering. Thus, the phenomenological boundary condition at the aggregate surface remains $c(0) = \beta$.

The authors gratefully acknowledge helpful discussions with Drs. H. L. Frisch, D. Mozyrsky and Y. Shnidman.